# ALBANK - A CASE STUDY ON THE USE OF ETHEREUM BLOCKCHAIN TECHNOLOGY AND SMART CONTRACTS FOR SECURE DECENTRALIZED BANK APPLICATION

**Shkëlqim SHERIFI\* [1], Shpend ISMAILI [1], Florim IDRIZI [1], Ejup RUSTEMI [1]**

[1] *\*Department of Computer Science, Faculty of Natural Sciences and Mathematics, RNM*
*\*Corresponding Authors: e-mail: s.sherifi211185@unite.edu.mk*

**Abstract**

New technologies, such as blockchain, are designed to address various system weaknesses, particularly those related to security. Blockchain can enhance numerous aspects of traditional banking systems by transforming them into digital, immutable, secure, and anonymous ledger.

This paper proposes a new banking application ALBank, which is based on blockchain and smart contract technologies. Its functionality relies on invoking functions within smart contracts deployed on the Ethereum blockchain. This approach enables decentralization and enhances both security and trust. In this context, the paper first presents a critical analysis of existing research on blockchain and traditional banking systems, with a focus on their respective challenges. It then examines the Know Your Customer (KYC) process and its various models. Finally, it introduces the design and development of ALBank, a decentralized banking application built on the Ethereum blockchain using smart contracts.

The results show that the integration of blockchain and smart contracts effectively addresses key issues in traditional banking systems, including centralization, inefficiency, and security vulnerabilities by storing critical data on a decentralized, immutable ledger, managing processes autonomously, and making transactions transparent to all users.

## 1. Introduction

Blockchain is a decentralized digital ledger that securely stores data on a network of computers [1], [21], [5], [6]. This technology has key features such as decentralization, integrity, immutability, verification, anonymity, and transparency. It allows participants in the blockchain network to write, read, and verify transactions recorded on a distributed ledger, while prohibiting the deletion or modification of existing transactions. Blockchain is secured by cryptographic primitives and protocols, such as encryption, decryption, digital signatures, hash functions, Transport Layer Security (TLS) protocol, Secure Shell (SSH) protocol, etc. These mechanisms ensure that transactions recorded on the ledger are authentic, verifiable, tamper-resistant, and non-repudiable. The distributed nature of blockchain requires all participants to maintain consensus on a unified version of the ledger. This is achieved through consensus protocols, sets of rules that all participants must follow to reach agreement and maintain global consistency [6].

The findings of this study are significant as they propose an innovative solution for a modern decentralized banking application (DAAP). The integration of blockchain and smart contracts can help the development of a more secure and reliable money transfer DAPP. This approach reduces the risk of security breaches, lowers transaction costs, ensures 24/7 availability, and enhances user trust.

This paper aims to address the following research questions:
- What are the existing problems in traditional banking systems?
- How can blockchain and smart contracts address these problems?

- What are the challenges involved in developing a blockchain-based decentralized bank application, and how can such an application be effectively designed and developed?

The remainder of this paper is organized as follows: **Section 1** provides a brief overview of blockchain technology, outlines the research questions addressed in this study, and describes the overall structure of the paper. **Section 2** discusses related work, focusing on contemporary implementations of blockchain and smart contracts aimed at addressing existing challenges in banking systems. **Section 3** introduces blockchain technology and smart contracts, including their historical development, along with the evolution of traditional banking systems, highlighting key challenges and limitations. **Section 4** explores the Know Your Customer (KYC) process. It begins with an overview of traditional KYC and its associated challenges, followed by a discussion of web-based KYC. It then examines non-blockchain-based KYC solutions, and finally introduces blockchain-based KYC. **Section 5** details the design and development of the blockchain-based DAPP - ALBank. **Section 6** presents performance metrics for the proposed DAAP, ALBank. The evaluation includes transaction speed, gas consumption, and network costs associated with executing transactions. Finally, **Section 7** concludes the paper by summarizing the key findings and contributions and outlining potential directions for future research.

A. *Methodology*

This paper is focused on addressing a specific problem: the development of a decentralized bank application based on blockchain and smart contract technologies.
A quantitative research methodology was employed. The developed application operates on the public Ethereum blockchain. Application efficiency and accuracy were evaluated using ALBank, through the analysis of ten iterations across various functions. The resulting numerical data are presented in tabular form and visualized through corresponding graphs.
Data collection was conducted using numerous reliable sources, including scientific journals, conference proceedings, and websites.

## 2. Literature review

In this section, we present papers in the field of proposed solutions for Blockchain-based Banking Systems and the KYC process. Based on this, we identified research challenges in the current work and proposed our system in this paper to address these challenges-ALBank [15].
A proposed solution introduces offline peer-to-peer schemes utilizing one-time programs (OTPs) executed within trusted execution environments (TEEs) [10]. While the system offers anonymity, programmability, and cost efficiency, it remains only partially decentralized due to the bank's role in issuing and signing OTPs—functionality that could be replaced by smart contracts.
Another study presents a blockchain-based Central Bank Digital Currency (CBDC) system that adopts a hybrid architecture, combining centralized and decentralized components [18]. The system employs Ethereum and smart contracts alongside a Proof-of-Authority (PoA) consensus algorithm to ensure scalability. Although digital assets are centrally controlled to comply with regulatory requirements, transactions are executed autonomously in a decentralized manner.
A separate approach utilizes a private blockchain to eliminate third-party intermediaries in interbank transactions, thereby reducing latency and enhancing security [13]. This system is built on Hyperledger Fabric and employs the Practical Byzantine Fault Tolerance (PBFT) consensus mechanism, emphasizing scalability and energy efficiency. BankTokens (BT) are introduced for cross-border exchanges, bypassing traditional currency conversion restrictions. However, node participation is restricted to authorized banks, ensuring centralized oversight.

A hybrid CBDC system has also been proposed to support confidential transactions while maintaining compliance with anti-money laundering and counter-terrorist financing regulations [8]. The architecture includes an account-based subsystem for transparency and a UTXO-based subsystem for privacy, with confidentiality ensured through ring signatures and zero-knowledge proofs. Despite these privacy features, the system remains partially centralized due to institutional monitoring.

Another contribution focuses on decentralizing the KYC process using smart contracts on the Ethereum blockchain [7]. This system streamlines data collection during loan applications, eliminates redundant verification steps, reduces operational costs, and enables real-time interbank data sharing. However, its application is limited to loan servicing.

Finally, a hybrid CBDC system is proposed that integrates UTXO-based and account-based subsystems to optimize transaction throughput [21]. The system introduces the POA-PBFT consensus algorithm, which is reported to be 50% more efficient than traditional PBFT. It prioritizes account-based transactions for high-frequency payments and UTXO for complex financial assets. However, the presence of a central authority conflicts with the principles of full decentralization and confidentiality.

## 3. Blockchain Technology, Smart Contracts, and Traditional Banking Systems

Chaum, in his doctoral thesis, was the first to propose a protocol conceptually similar to blockchain [5]. In 1991, Haber and Stornetta introduced a cryptographically secured chain of blocks. In 1998, the Bit Gold project proposed a decentralized digital currency mechanism [14]. In 2008, Nakamoto introduced Bitcoin, a fully peer-to-peer electronic cash system [11], and the term blockchain was first used to describe the distributed ledger underlying Bitcoin transactions [16]. In 2013, Buterin proposed Ethereum in his white paper [3].

According to Nakamoto, blockchain is a distributed ledger system that enables the creation of secure and transparent digital records [11]. It operates through a decentralized network of nodes that verify and validate transactions to ensure consistency and immutability [9]. Blockchain is characterized by peer-to-peer communication, eliminating the need for third-party intermediaries. Figure 1 illustrates the connection of blocks within the blockchain.

The term smart contracts were first introduced in 1994 by computer scientist and cryptographer Nick Szabo [17], who defined a smart contract as: "A smart contract is a computerized transaction protocol that executes the terms of a contract" [17]. A smart contract is activated by sending a transaction to it, which triggers its autonomous execution across all nodes in the network, based on the data contained in the initiating transaction.

The history of banking dates back to around 2000 BC in ancient Assyria, India, and Sumer [20]. The establishment of the Bank of Amsterdam in 1609 marked a significant milestone in the evolution of modern banking, as it was the first institution to issue banknotes of value and offer safekeeping services for valuable assets [4]. In the late 19th and early 20th centuries, a series of financial crises prompted major reforms in the U.S. banking system [19]. One of the most notable developments was the enactment of the National Bank Act in 1863, which introduced a standardized national currency and laid the foundation for federal regulation of banks [19].

Blockchain technology can establish a new paradigm for transferring valuable assets, supporting business operations, enhancing transparency, and fostering trust. According to the Bank for International Settlements (BIS), the application of Distributed Ledger Technology (DLT) could fundamentally transform how assets are stored and maintained, obligations are fulfilled, contracts are enforced, and risks are managed [2].

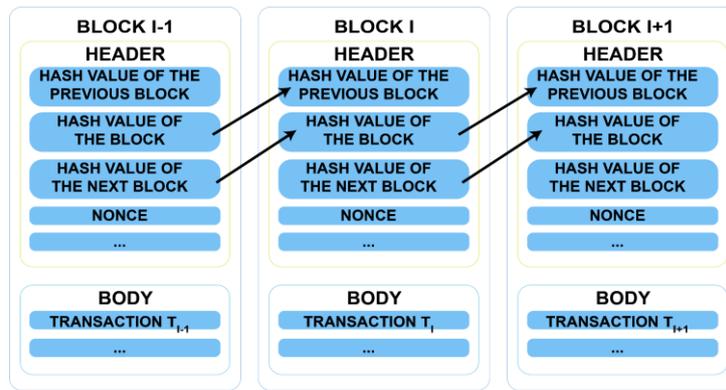

*Figure 1.* Block Connections in a Blockchain

### A. Problems in the Traditional Banking System

*1) Transaction Cost and Speed*: Traditional banking involves high transaction fees and slow processing, often requiring physical presence.
2). *Centralized Control*: Centralized banking lacks user anonymity and poses risks to data privacy and security.
3). *Financial Crises*: Loan defaults in centralized systems can trigger widespread financial instability and economic downturns.
4). *Security Threats*: Traditional banks are vulnerable to various attacks that compromise availability, confidentiality, integrity, and authenticity.

## 4. Process Know Your Customer (KYC)

The Know Your Customer (KYC) process is a regulatory requirement rooted in the Money Laundering Control Act of 1986 [12]. It mandates that financial institutions verify the identity of their clients to comply with legal standards.

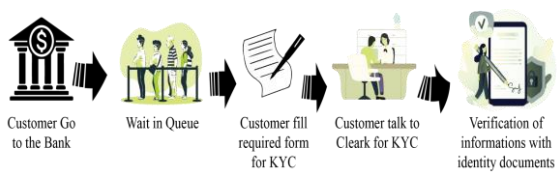

*Figure 2.* Traditional KYC Process

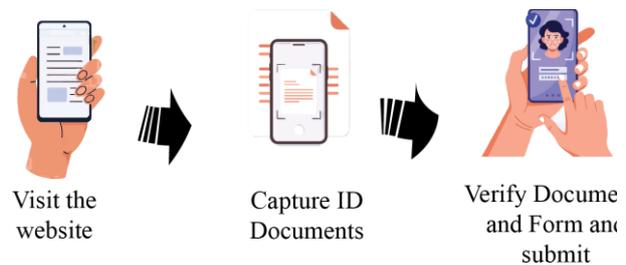

*Figure 3.* Web-Based KYC Process

The traditional KYC process is manual, time-consuming, and resource-intensive. It typically requires customers to be physically present at a bank, complete paperwork, and exchange sensitive documents with bank clerks. This approach not only includes high operational costs but also compromises confidentiality and user experience. Figure 2 illustrates the traditional KYC process. In contrast, the web-based KYC process leverages digital platforms to streamline identity verification. Customers can complete the process remotely by uploading scanned documents and submitting electronic forms. This method significantly reduces time, cost, and human error while enhancing confidentiality and convenience. Figure 3 illustrates the web-based KYC process.

While web-based KYC improves efficiency over traditional methods, it still relies on centralized databases, which remain vulnerable to breaches and lack interoperability across institutions.

The non-blockchain-based KYC process, whether traditional or web-based, becomes increasingly inefficient when a customer interacts with multiple financial institutions. Each institution must independently repeat the entire KYC process, leading to redundancy, increased costs, and security risks. In contrast, the blockchain-based KYC process introduces a decentralized, secure, and transparent alternative. Customer data is stored on a distributed ledger, accessible to authorized institutions without the need for repeated verification. This model enhances data integrity, reduces operational overhead, and ensures compliance with regulatory standards. Public blockchains like Ethereum can be used to implement such systems, enabling real-time, tamper-proof access to verified identities.

## 5. Blockchain-Based Decentralized Bank Application Design and Development - ALBank

With the increasing use of personal computers and smart devices, along with the rapid development of e-business, new challenges have emerged concerning data security, integrity, efficiency, and transaction speed, particularly within institutions responsible for their execution, such as traditional banking systems [13]. Table 1 presents a comparative analysis between traditional banking systems and blockchain-based banking systems.

*Table* 1 . Comparison of Traditional Banking System and Blockchain Banking System

| Parameters | Traditional Banking System | Banking System on Blockchain |
|---|---|---|
| **Transaction Speed** | Slower due to intermediaries | Faster due to peer-to-peer transactions |
| **Transaction Cost** | Higher due to fees | Lower due to reduced intermediaries |
| **Transparency** | Limited visibility into transaction processes | Full transparency with Blockchain technology |
| **Security** | Centralized security measures | Decentralized and cryptographic security |
| **Privacy** | Personal data is stored and managed by the bank | Enhanced privacy with pseudonymous transactions |
| **Access** | Limited by banking hours and geographic location | Accessible 24/7 from anywhere with internet |
| **Scalability** | Limited by infrastructure | High scalability with Blockchain technology |
| **Control** | Limited Control over Funds | Full Control over Funds |
| **Regulatory** | Strict regulations and compliance requirements | Generally less strict |
| **Innovation** | Slower adoption of new technologies | Rapid innovation and integration of technologies |

The main challenges in traditional banking systems are related to data security and operational efficiency. As data is stored on distributed servers, there is a heightened risk of unauthorized access, data manipulation, and cyberattacks. These issues necessitate advanced technological solutions to ensure the security and reliability of stored data.

In recent years, financial institutions and banks have increasingly adopted blockchain technology due to its numerous advantages [8]. The decentralized architecture of blockchain enhances transaction speed and system management by eliminating the need for intermediaries and reducing transaction costs. Its distributed logic increases usability and flexibility, enabling routine and autonomous execution of operations across the network. This autonomous execution, facilitated by smart contracts, significantly improves efficiency and optimizes operational workflows.

To address these challenges and leverage the advantages of blockchain, we propose ALBank, a decentralized, blockchain-based application. ALBank is developed using different programming languages and libraries. The front end is built with HTML and CSS. Functionality is supported by pure JavaScript, and the Bootstrap library is used to provide a consistent and standard UI/UX. Interaction between the front end and the back end (the smart contract) is facilitated by the Node.js Server and Web3 library. MetaMask is used for user authentication. The back end of ALBank is implemented by writing smart contracts in Solidity without external libraries. ALBank is tested on the Sepolia testnet network. The public Ethereum blockchain is used to deploy the contract and power ALBank.

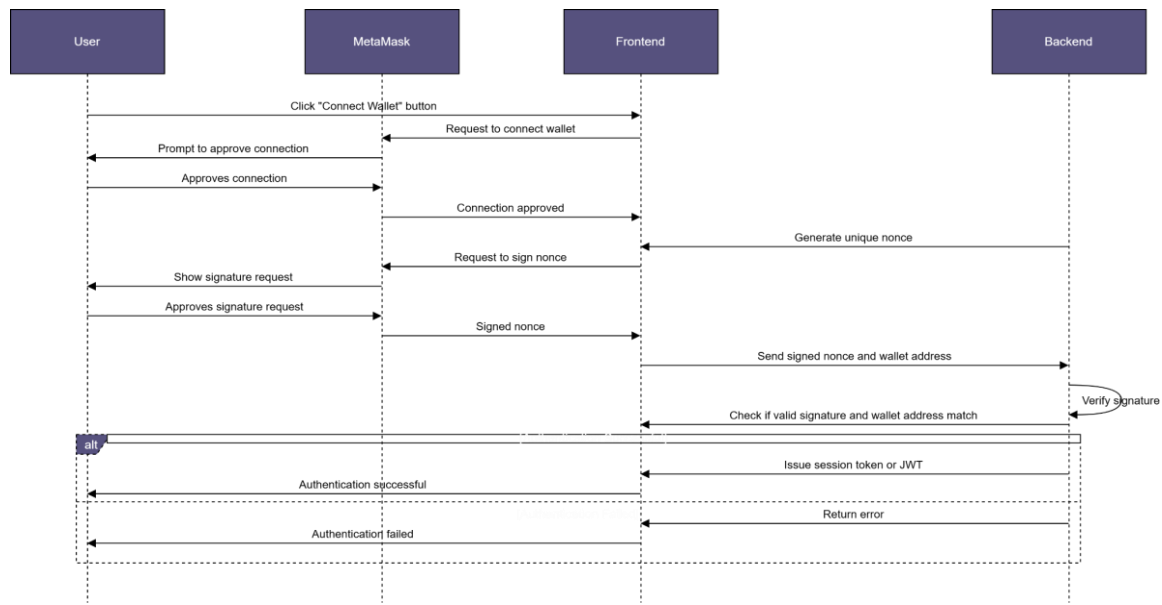

*Figure 4.* Sequence Diagram of Authentication and Login – ALBank.

Authentication on ALBank is processed using MetaMask, which allows users to authenticate themselves without a traditional username and password Figure 4. The process follows these steps:
1. *User clicks "Connect Wallet"* - Initiates the authentication process.
2. *Frontend sends a connection request to MetaMask* - MetaMask prompts the user.
3. *User approves the connection in MetaMask* - MetaMask confirms the connection to the frontend.
4. *Frontend requests MetaMask to sign a nonce* - A unique, one-time-use number is used to prevent replay attacks.
5. *MetaMask prompts the user to sign the nonce* - User approves the signature request.

6. *MetaMask returns the signed nonce to the frontend* - Frontend sends the signed nonce and wallet address to the backend.
7. *Backend verifies the signature* - Checks if the signature is valid. Confirms it matches the wallet address.
8. *Return:* If valid: Backend issues a session token or JWT. If invalid: Backend returns an authentication error.

```html
<div class="image-container">
  
  <div class="overlay-content">
   <h1>ALBank</h1>
   <button onclick="loginWithMetaMask()">Login with MetaMask</button>
   <div class="status" id="status">Connect your wallet to continue</div>
  </div>
</div>
<script>
    loginButton.addEventListener('click', async () => {
      var userAddress = document.getElementById('accountaddress');
      var accounts = await ethereum.request({ method: 'eth_requestAccounts'})
      address = accounts[0];
      userAddress.innerText = address;
      userAddress.classList.remove('d-none');
      loginButton.classList.add('d-none');
      console.log(accounts);
      console.log(accounts[0]); });
      ethereum.on('accountsChanged', async function (accounts) {
      var userAddress = document.getElementById('accountaddress');
      var accounts = await ethereum.request({ method: 'eth_requestAccounts'})
      address = accounts[0];
      userAddress.innerText = address; });
</script>
```

*Figure 5.* Part of code for login ALBank.

The code sequence in Figure 5 shows a simple HTML login form on ALBank, where a button is defined with an onclick event that triggers the loginWithMetaMask function. The code also includes the implementation of the loginWithMetaMask function, which asynchronously retrieves the user's account address. It then sends a request to Ethereum using the Web3 library with the requested address, prompting the MetaMask login form. After successful authentication, the user is granted access to the ALBank home page.

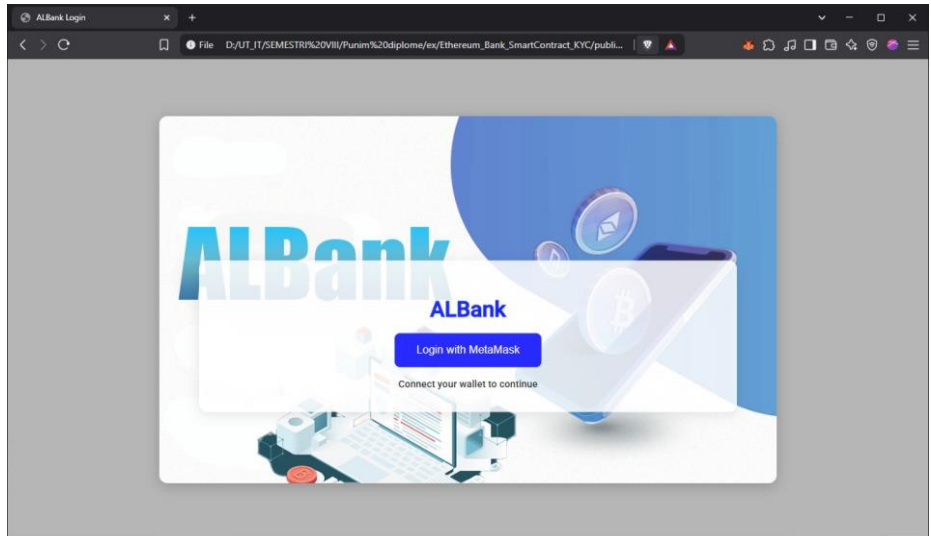

*Figure 6.* Home Login page - ALBank.

The login page user interface on ALBank is simple and intuitive Figure 6. The page contains a single button that triggers an event to open the MetaMask login form. Authentication is then handled through MetaMask, after which the user is redirected to the ALBank home page.

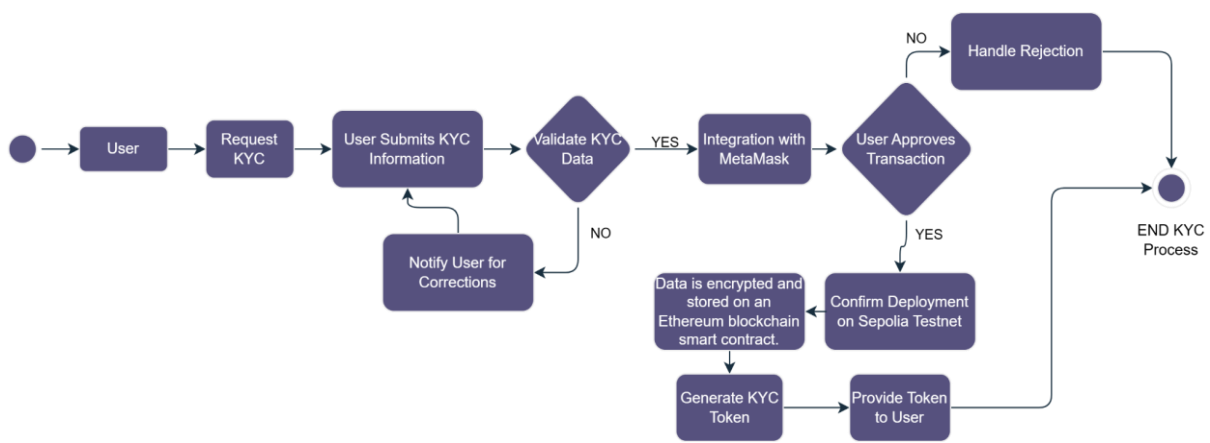

*Figure 7.* Flowchart Diagram of KYC – ALBank.

KYC on ALBank allows users to submit personal information and securely store it on the Ethereum blockchain illustrated in Figure 7. Once the KYC data is recorded on the blockchain, a transaction address is generated. This address can be used to retrieve the user's information without requiring them to re-enter their details when opening another bank account. The process follows these steps:
1. *User* - The process begins with the user initiating the KYC procedure.
2. *Request KYC* - The module sends a request to the user to begin the KYC process.
3. *User Submits KYC Information* - The user provides the required identity and verification details.
4. *Validate KYC Data* - Checks the submitted data for accuracy and completeness. If validation fails: The user is notified to correct the data. If validation succeeds: Proceed to MetaMask integration.
5. *Integration with MetaMask* - Prepares a blockchain transaction.
6. *User Approves Transaction* - MetaMask prompts the user to approve the transaction. If the user rejects: The KYC process ends. If the user approves: Proceed to deployment.

7. *Confirm Deployment on Sepolia Testnet* - The validated KYC data is deployed to a smart contract on the Sepolia Ethereum testnet.
8. *Data is Encrypted and Stored* - The KYC data is securely encrypted and stored on the Ethereum blockchain.
9. *Generate KYC Token* - A unique token representing the user's verified identity is created.
10. *Provide Token to User* - The token is issued to the user as proof of successful KYC.

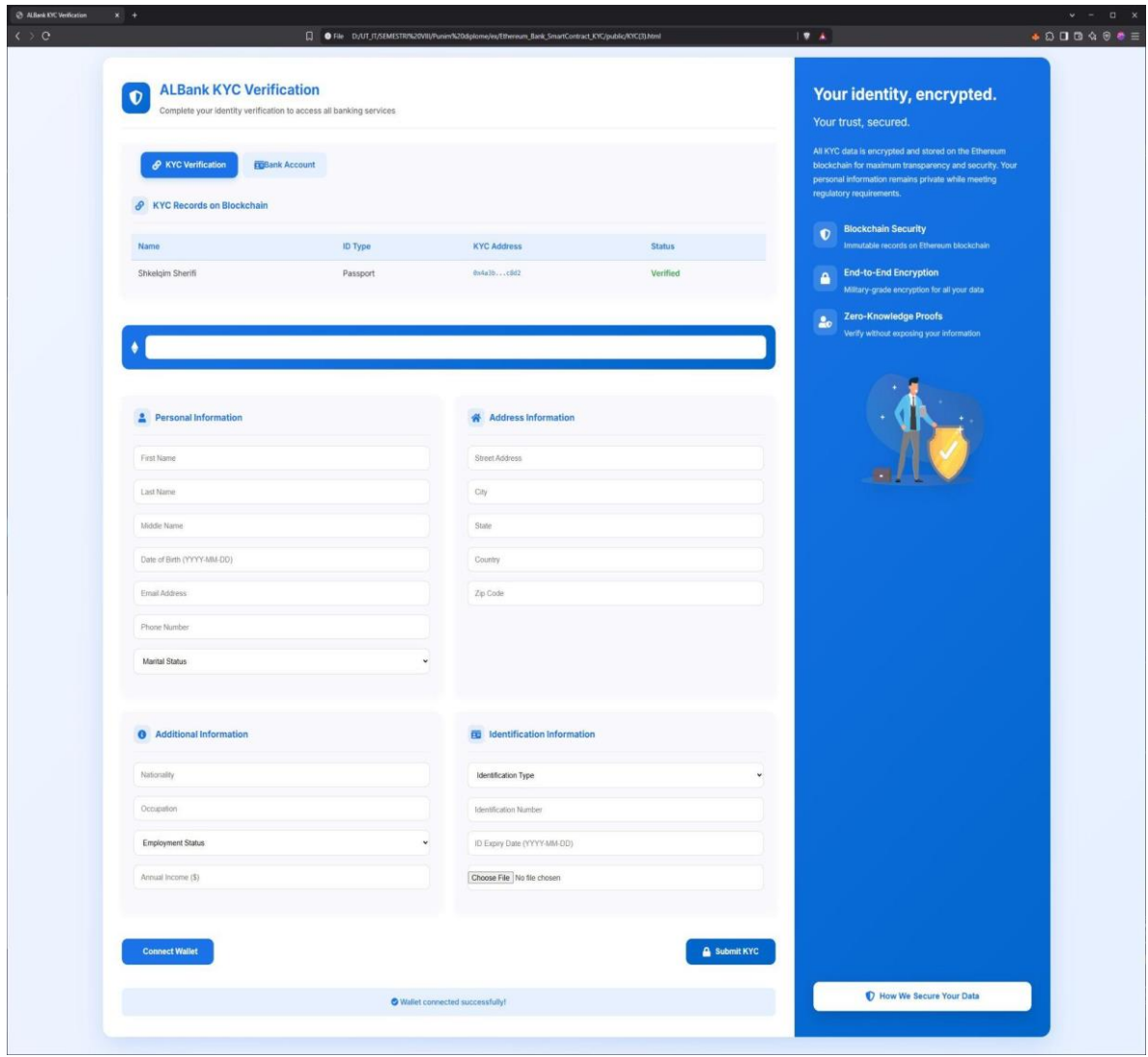

*Figure 8.* KYC page - ALBank.

The KYC page user interface on ALBank is simple, intuitive, and follows the Material Design theme Figure 8. The page contains a table that lists registered KYC records on the blockchain along with their corresponding transaction addresses and a form with the following input groups:
1. *Personal Information* – collects basic personal details about the user.
   The group in the form consists of personal information fields with the following inputs:
      a. *First Name* – type="text". Collects the user's first name.
      b. *Last Name* – type="text". Collects the user's last name.
      c. *Middle Name* – type="text". An optional field for the user's middle name.
      d. *Date of Birth* – type="text". Collects the user's birth date in YYYY-MM-DD format.

e. *Email Address* – type="email". Collects the user's email for communication and verification.
  f. *Phone Number* – type="tel". Collects the user's phone number.
  g. *Marital Status* – select dropdown. Options: Single, Married, Divorced, Widowed. Collects the user's marital status.
2. *Address Information* – collects the user's residential address details.
  The group in the form consists of address information fields with the following inputs:
  a. *Street Address* – type="text". Collects the user's street address.
  b. *City* – type="text". Collects the user's city.
  c. *State* – type="text". Collects the user's state or province.
  d. *Country* – type="text". Collects the user's country.
  e. *Zip Code* – type="text". Collects the user's postal/zip code.
3. *Additional Information* – collects more specific personal and professional details about the user.
  The group in the form consists of additional information fields with the following inputs:
  a. *Nationality* – type="text". Collects the user's nationality.
  b. *Occupation* – type="text". Collects the user's job title or role.
  c. *Employment Status* – select dropdown. Options: Employed, Self-Employed, Unemployed, Student, Retired. Collects the user's current employment status.
  d. *Annual Income* – type="number". Collects the user's yearly income in USD.
4. *Identification Information* – collects data required for verifying the user's identity.
  a. *Identification Type* – select dropdown. Options: Passport, Driver's License, National ID. Let the user choose the type of ID they are submitting.
  b. *Identification Number* – type="text". Collects the ID number from the selected document.
  c. *ID Expiry Date* – type="text". Collects the expiration date of the ID in YYYY-MM-DD format.
  d. *Upload ID File* – type="file". Allows the user to upload a scanned copy or photo of their ID.

At the end of the form, the following buttons are available:
1. *Connect Wallet* – button. Triggers wallet connection (MetaMask).
2. *Submit KYC* – button. Submits the form data for verification and blockchain storage.

```solidity
pragma solidity ^0.8.0;
contract CombinedBankWithKYC {
    address public owner;
    struct UserRegistrationData {
        string firstName; string middleName; string lastName; string dob; string email; string phone; string maritalStatus; string address_; string city; string state; string country; string zip; string nationality; string occupation; string employmentStatus; uint annualIncome; string idType; string idNumber; string idExpiry;
    }

    function registerUser(UserRegistrationData memory data) public onlyOnce {
        require(bytes(data.firstName).length > 0, "First name is required");
        require(bytes(data.lastName).length > 0, "Last name is required");
        require(bytes(data.dob).length > 0, "Date of birth is required");
        require(bytes(data.email).length > 0, "Email is required");
        require(bytes(data.phone).length > 0, "Phone number is required");
        require(bytes(data.address_).length > 0, "Address is required");
        require(bytes(data.city).length > 0, "City is required");
        require(bytes(data.state).length > 0, "State is required");
        require(bytes(data.country).length > 0, "Country is required");
        require(bytes(data.zip).length > 0, "ZIP code is required");
        require(bytes(data.idType).length > 0, "ID type is required");
        require(bytes(data.idNumber).length > 0, "ID number is required");
        users[msg.sender] = User(
            data.firstName, data.middleName, data.lastName, data.dob, data.email, data.phone, data.maritalStatus, data.address_, data.city, data.state, data.country, data.zip, data.nationality, data.occupation, data.employmentStatus, data.annualIncome, data.idType, data.idNumber, data.idExpiry
        );
        emit UserRegistered(msg.sender);
    }
    function getUser(address _userAddress) public view returns (User memory) {
        require(_userAddress != address(0), "Invalid address");
        require(bytes(users[_userAddress].firstName).length > 0 && bytes(users[msg.sender].idType).length == 0 && bytes(users[msg.sender].idNumber).length == 0, "User does not exist");
        return users[_userAddress];
    }
    modifier onlyOnce() {
        require(bytes(users[msg.sender].firstName).length == 0 && bytes(users[msg.sender].idType).length == 0 && bytes(users[msg.sender].idNumber).length == 0, "User is already registered"); _;}
```

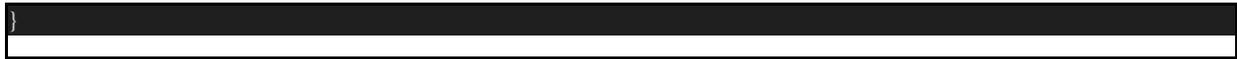
}

*Figure 9.* Part of the Solidity smart contract code for registering and retrieving user KYC data for ALBank.

The code section in Figure 9. shows the implementation of the smart contract CombinedBankWithKYC. The contract begins by defining a variable of type address to store the owner's address. A structure named UserRegistrationData is defined with the following fields:
string firstName; string middleName; string lastName; string dob; string email; string phone; string maritalStatus; string address_; string city; string state; string country; string zip; string nationality; string occupation; string employmentStatus; uint annualIncome; string idType; string idNumber; string idExpiry;.
These fields map the input data from the KYC form page.
Next, the registerUser public function is defined. It takes a UserRegistrationData structure as input, passed from memory. This function is marked with the onlyOnce modifier, which ensures that no duplicate data is stored on the blockchain by checking the user's first name, ID type, and ID number. This guarantees that each user's KYC data is stored only once. All structure fields are validated to ensure they are properly filled and not left blank. Once validation passes, the user is registered on the blockchain, and the UserRegistered event is emitted with the user's address as a parameter.
The getUser function takes an address as a parameter and retrieves the corresponding KYC data. Before returning the data, the function checks whether the address is valid and whether a user with the given first name, ID type, and ID number exists. This function is public, marked as view, and returns the User structure stored in temporary memory.
The flowchart of the Bank DApp – ALBank Figure 10. outlines the sequence of user activities and the logic for interacting with the decentralized banking platform managed by a smart contract. First, the user needs to be authenticated through MetaMask. The authentication process is described in Figure 4. Once authentication is complete, the user can access the ALBank DApp. Next, the user must complete the KYC process, as explained in Figure 7. After successful KYC verification, the user can perform various banking actions directly in a decentralized manner using the smart contract.
The process follows these steps:
1. *User* - The process begins with the user initiating interaction.
2. *Authentication Module* - The user is authenticated to ensure secure access.
3. *Access the DApp (ALBank)* - Upon successful authentication, the user gains access to the ALBank decentralized application.
4. *DAAP - ALBank* - The user interface of ALBank is presented for interaction. Decision Point: "Does the user have KYC registered on the blockchain?". If NO → The user is directed to the KYC Module to complete their Know Your Customer registration. If YES → The user proceeds to Banking Operations.
5. *KYC Module* - This module handles the registration of user identity and verification data on the blockchain.
6. *Banking Operations* - Once KYC is verified, the user can perform various banking functions:
    a. Add Customer
    b. Add KYC Customer Data
    c. Get KYC Data
    d. Deposit ETH
    e. Withdraw ETH
    f. Get Bank Balance

7. *Smart Contract* - All operations are executed through smart contracts deployed on the Ethereum blockchain, ensuring transparency and security.

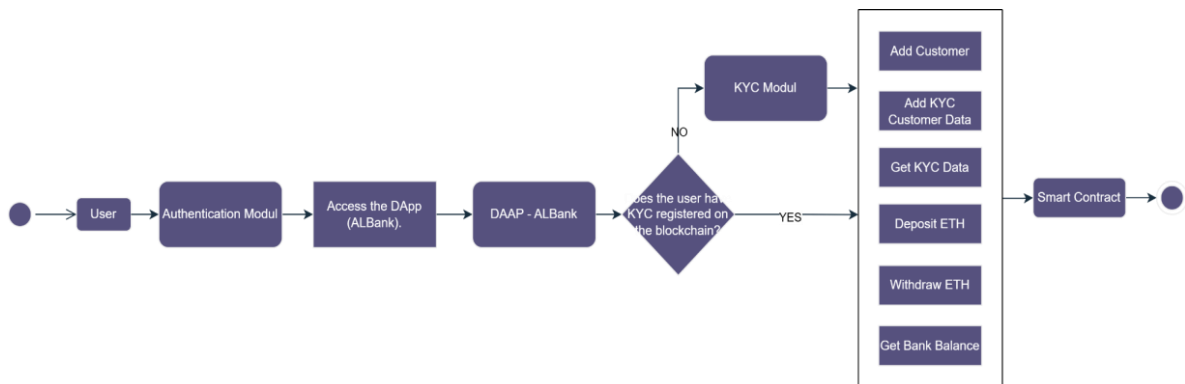

*Figure 10.* Flowchart Diagram of Bank DAPP – ALBank.

The Bank Account page user interface on ALBank follows the overall design theme, Figure 11. It is simple, intuitive, and uses the Material Design style. At the top of the page, there is an input field where the user can enter the address of their account.

The Bank Account Operations form consists of three main groups:
1. *Deposit ETH Group*:
    a. *Deposit Amount (ETH)* – type="number": Collects the amount of ETH the user wants to transfer to the smart contract.
    b. *Deposit Button*: Triggers the transfer of ETH from the user's account to the smart contract.
2. *Withdraw ETH Group:*
    a. *Withdraw Amount (ETH)* – type="number": Collects the amount of ETH the user wants to withdraw from the smart contract to their account.
    b. *Withdraw Button*: Triggers the transfer of ETH from the smart contract to the user's account.
3. *Bank Balance Group:*
    a. *Bank Balance Textbox* – type="text": Displays the amount of ETH linked to the smart contract.
    b. *Refresh Balance Button:* Triggers a refresh of the displayed ETH amount linked to the smart contract.

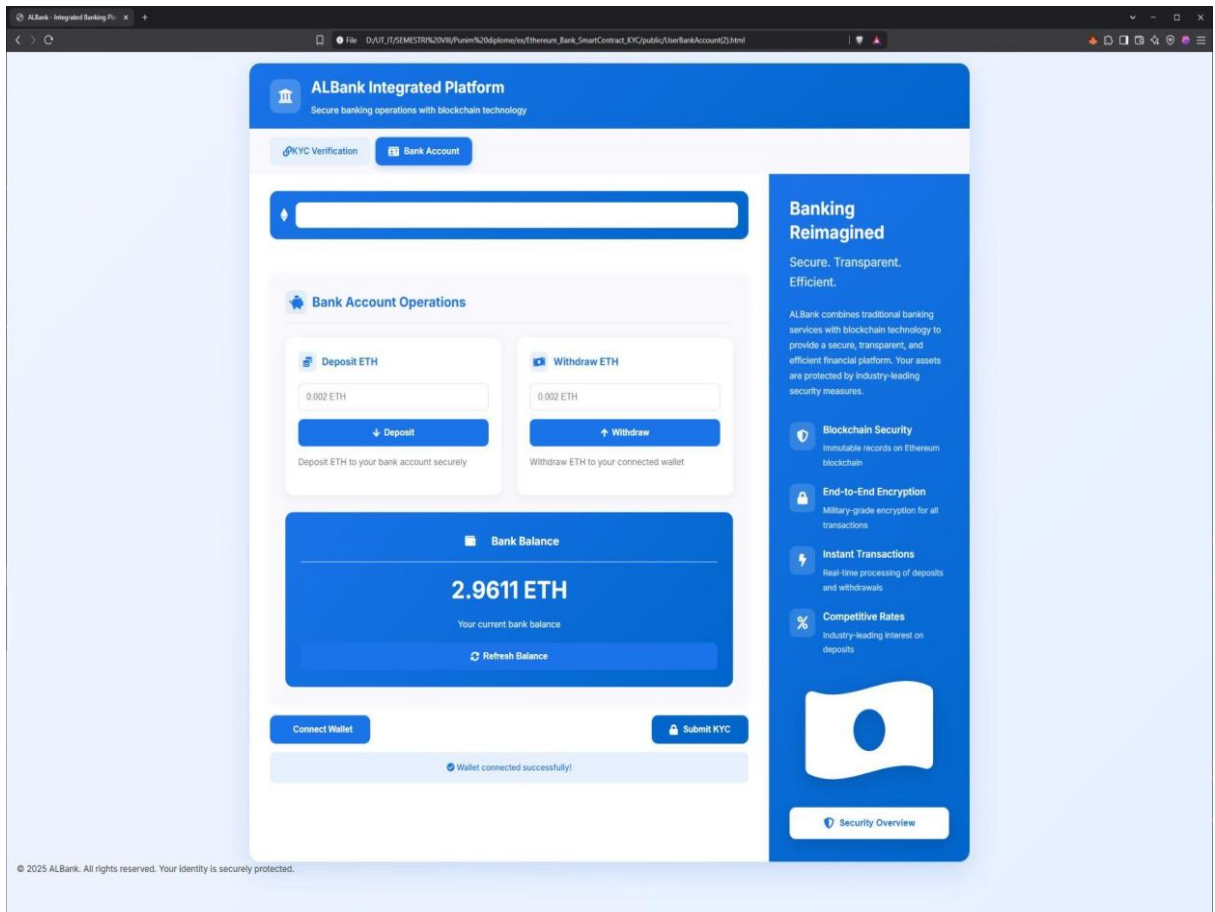

*Figure 11.* Bank page - ALBank.

```
document.addEventListener('DOMContentLoaded', async () => {
...
    depositButton.addEventListener('click', () => {
      console.log("Started executing depositButton");
      var start = performance.now();
      var depositInput = document.getElementById('depositeth');
      myContract.methods.deposit().send({ from: address, value: web3.utils.toWei(depositInput.value, 'ether') })
        .then(res => {...});
      })
        .catch(err => {...});});
    withdrawButton.addEventListener('click', () => {
      console.log("Started executing withdrawButton");
      var start = performance.now();
      var withdrawInput = document.getElementById('withdraweth');
      myContract.methods.withdraw(web3.utils.toWei(withdrawInput.value, 'ether')).send({ from: address })
        .then(res => {...})
        .catch(err => {...}); });
```

```javascript
    getBalanceButton.addEventListener('click', () => {

      console.log("Started executing getBalanceButton");

      var start = performance.now();
      myContract.methods.getbalance().call({ from: address })

        .then(res => {...});

      })

      .catch(err => {...});});
}
```

*Figure 12.* Part of the JavaScript code for performing deposit, withdrawal and get balance actions on ALBank.

The code section in Figure 12. shows the implementation of the *depositButton*, *withdrawButton*, and *getBalanceButton* click events. Each function measures the elapsed time by recording the start and end times in a variable.
1. *depositButton*: This function first retrieves the input element with the html id "*depositeth*", then calls the *deposit*() method defined in the smart contract. The *send* () function is used to send the transaction, passing the user's address and the input value converted to Ether.
2. *withdrawButton*: This function retrieves the input element with the html id "*withdraweth*", then calls the *withdraw*() method with the input value converted to Ether. The transaction is sent using the *send* () function with the user's address.
3. *getBalanceButton*: This function calls the *getBalance*() method, which uses the *call*() function to retrieve the balance associated with the user's address.

```solidity
function deposit() public payable returns (bool) {

  uint256 startGas = gasleft();

  uint256 startTime = block.timestamp;

  require(msg.value > 10 wei, "Please deposit at least 10 wei");

  userbalance[msg.sender] += msg.value;

  emit Deposit(msg.sender, msg.value);

  uint256 gasUsed = startGas - gasleft();

  uint256 timeTaken = block.timestamp - startTime;

  emit GasConsumption(gasUsed);

  emit ElapsedTime(timeTaken);

  return true;

}
function withdraw(uint256 _amount) public nonReentrant payable returns (bool) {

  uint256 startGas = gasleft();

  uint256 startTime = block.timestamp;

  require(_amount <= userbalance[msg.sender], "You do not have sufficient balance");

  userbalance[msg.sender] -= _amount;
```

```solidity
        payable(msg.sender).transfer(_amount);
        emit Withdrawal(msg.sender, _amount);
        uint256 gasUsed = startGas - gasleft();
        uint256 timeTaken = block.timestamp - startTime;
        emit GasConsumption(gasUsed);
        emit ElapsedTime(timeTaken);
        return true;
    }
    function getbalance() public view returns (uint256) {
        return userbalance[msg.sender];
    }
    modifier nonReentrant() {
        require(!locked, "Reentrant call");
        locked = true; _; locked = false;
    }
```

*Figure 13.* Part of the Solidity code for performing deposit, withdrawal and get balance actions on ALBank.

The code section in Figure 13 shows the implementation of the *depositButton*, *withdrawButton*, and *getBalanceButton* functions in the smart contract using Solidity. Each function measures the elapsed time and gas used for the transaction for log.
1. *deposit*: This function first checks that the minimum value sent is greater than 10 wei. If the condition is met, the amount is added to the user's balance. Then, a Deposit event is emitted with the sender's address and the deposited value. This function is marked with the payable modifier, which allows it to receive Ether.
2. *withdraw*: This function first checks that the user's balance is greater than or equal to the requested withdrawal amount. If the condition is met, the amount is deducted from the user's balance, and the transfer function is called to send Ether to the sender's address. A Withdraw event is then emitted with the sender's address and the withdrawn amount. This function is marked with the payable and nonReentrant modifiers to allow Ether transfers and to prevent reentrancy attacks or multiple simultaneous calls.
3. *getBalance*: This function returns the balance of the user based on their address.

## 6. Performance Metrics

After completing the development of the Alpha-2 version of the ALBank, testing was conducted, and the results are presented in Table II. For each function, 10 samples were collected. Repeating the test 10 times provides a more representative and reliable assessment of the application's performance. Three parameters were evaluated for each function: transaction speed, gas consumption for transaction completion, and network fee. The testing focused on the core functionalities of the application, which include: registering a new customer, submitting KYC data for a new customer, retrieving KYC data for an existing customer, depositing ETH, withdrawing ETH, and retrieving the account balance.
Figure 14. and Figure 16. illustrates the average transaction speed. As shown in the graph, the average execution time for data retrieval functions is nearly zero. In contrast, the functions

related to data insertion range between $1 * 10^4$ ms to $2 * 10^4$ ms. The function with the highest average execution time is the one responsible for adding KYC data for a new customer, which exceeds $4 * 10^4$ ms. As shown in the graph, data retrieval functions execute rapidly, typically within a few milliseconds. Functions that write data to the blockchain generally require between one and three seconds to complete. Notably, the function responsible for adding KYC data for new customers is both prominent and resource-intensive, with an associated cost approaching 0.030 ETH.

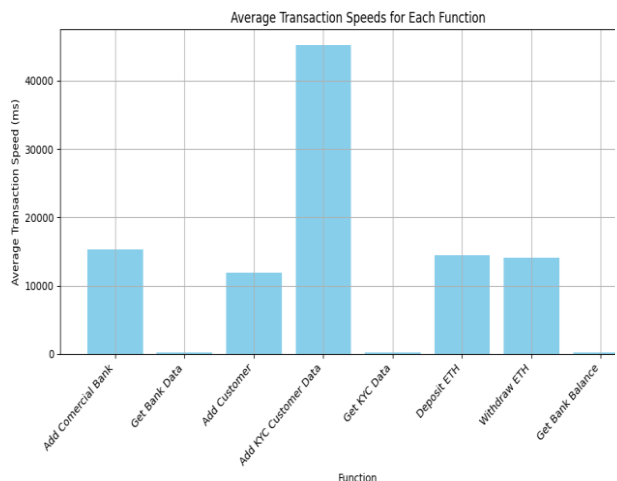

*Figure 14.* Average Transaction Speed of Functions.

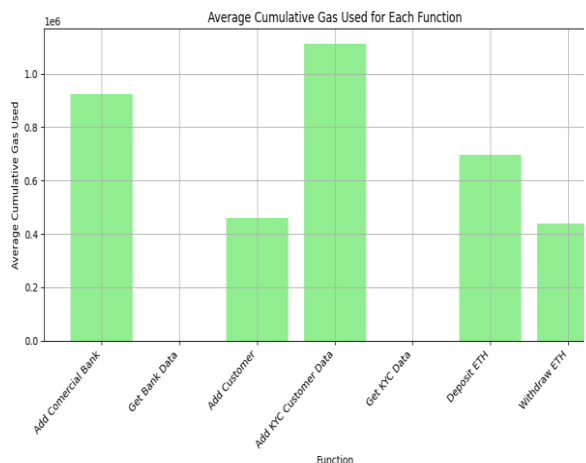

*Figure 15.* Average Gas Usage of Functions.

Figure 15. and Figure 17. illustrates the gas consumption required to complete each transaction. As depicted, gas usage varies by function, but in most cases does not exceed $2 * 10^5$ units. This reflects the computational effort required to execute each transaction.

The network fee for most functions remains minimal. However, the function for adding KYC data for a new customer is comparatively expensive, with a fee that can reach up to 0.040 ETH. From these observations, we conclude that this function is both time-consuming and costly relative to others in the DAPP.

Nevertheless, the use of blockchain ensures that KYC data is recorded once per user and can be reused by any application when opening new accounts. This significantly enhances the efficiency, speed, and cost-effectiveness of the KYC process compared to traditional methods.

*Table 2.* Ethereum Bank Smart Contract Performance Metrics

| Functions | Parameters | 1 | 2 | 3 | 4 | 5 | 6 | 7 | 8 | 9 | 10 |
|---|---|---|---|---|---|---|---|---|---|---|---|
| Add Customer | Transaction Speed | 12945.00 | 17552.10 | 10853.50 | 17268.40 | 8840.40 | 7211.40 | 8491.60 | 9356.20 | 16974.40 | 9071.30 |
| | Cumulative Gas Used | 269777.00 | 786207.00 | 183208.00 | 120208.00 | 830029.00 | 879395.00 | 248813.00 | 101221000 | 162208.00 | 12022000 |
| | Network fee | 0.00030 | 0.00030 | 0.00030 | 0.00030 | 0.00030 | 0.00030 | 0.00030 | 0.00030 | 0.00030 | 0.00030 |
| Add KYC Customer Data | Transaction Speed | 55304.70 | 7434.60 | 23645.30 | 54391.12 | 9345.42 | 34519.00 | 72345.37 | 82633.38 | 99412.72 | 13045.30 |
| | Cumulative Gas Used | 1608905.00 | 908905.00 | 1128905.00 | 908602.00 | 808712.00 | 1938612.00 | 538205.00 | 438601.00 | 1328603.00 | 1518202.00 |
| | Network fee | 0.00130 | 0.03850 | 0.03150 | 0.03280 | 0.02150 | 0.03270 | 0.03270 | 0.03270 | 0.03350 | 0.03830 |
| Get KYC Data | Transaction Speed | 171.00 | 706.20 | 42.10 | 307.00 | 28.60 | 316.90 | 24.40 | 18.20 | 321.10 | 12.10 |
| | Cumulative Gas Used Network | 0.0 | 0.0 | 0.0 | 0.0 | 0.0 | 0.0 | 0.0 | 0.0 | 0.0 | 0.0 |
| | fee | 0 0 0 0 | 0 0 0 0 | 0 0 0 0 | 0 0 0 0 | 0 0 0 0 | 0 0 0 0 | 0 0 0 0 | 0 0 0 0 | 0 0 0 0 | 0 0 0 0 |
| Deposit ETH | Transaction Speed | 21674.50 | 13350.60 | 17881.40 | 9484.60 | 15814.30 | 12106.90 | 17824.30 | 10413.80 | 10907.70 | 14659.80 |
| | Cumulative Gas Used Network fee | 69206.00 0.00010 | 73106.00 0.00010 | 796358.00 0.00010 | 211976.00 0.00010 | 140958.00 0.00010 | 243711.00 0.00010 | 4972421.00 0.00010 | 181662.00 0.00010 | 188156.00 0.00010 | 73106.00 0.00010 |
| Withdraw ETH | Transaction Speed | 14799.70 | 13103.50 | 14176.60 | 19066.90 | 11552.00 | 16263.10 | 19091.80 | 14302.70 | 5653.90 | 12432.30 |
| | Cumulative Gas Used Network fee | 565010.00 0.00020 | 1816584.00 0.00020 | 112000.00 0.00020 | 133000.00 0.00020 | 213391.00 0.00020 | 206163.00 0.00020 | 253994.00 0.00020 | 91000000.00 0.00020 | 167059.00 0.00020 | 84485.00 0.00020 |
| Get Bank Balance | Transaction Speed | 626.30 | 92.60 | 193.60 | 316.60 | 28.90 | 341.90 | 27.00 | 11.40 | 342.90 | 29.70 |
| | Cumulative Gas Used Network fee | 0.000000 | 0.000000 | 0.000000 | 0.000000 | 0.000000 | 0.000000 | 0.000000 | 0.000000 | 0.000000 | 0.000000 |

**Measurable units:**

Transaksionit Speed Unit - **ms**; Cumulative Gas Used - **unit(n)**; Network fee - **Sepolia ETH**

**Computer Specifications:**

CPU: Intel(R) Core(TM) i7-6700HQ CPU @ 2.60GHz; RAM: 32.0 GB; OS: Windows 10 Pro 22H2; Internet: A1 FiberOptic up to 200Mb/s; Network: Sepolia Test Net with Ethereum Blockchain

**Project URL alpha-version: https://github.com/shkelqimsherifi/Ethereum Bank SmartContract KYC**

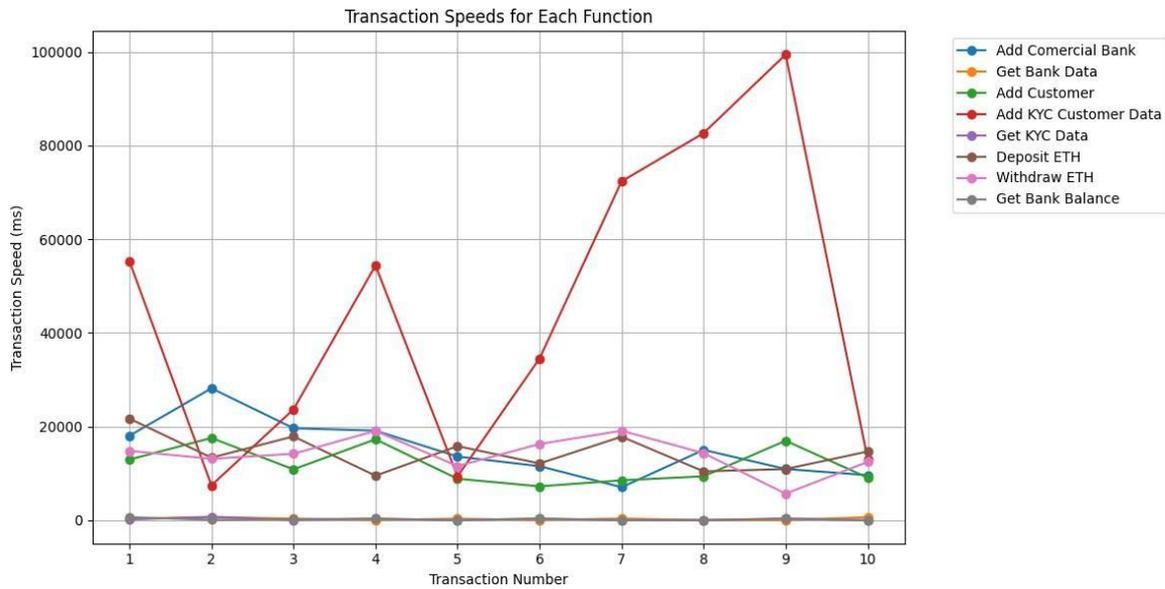

*Figure 16.* Transaction Speed of Functions.

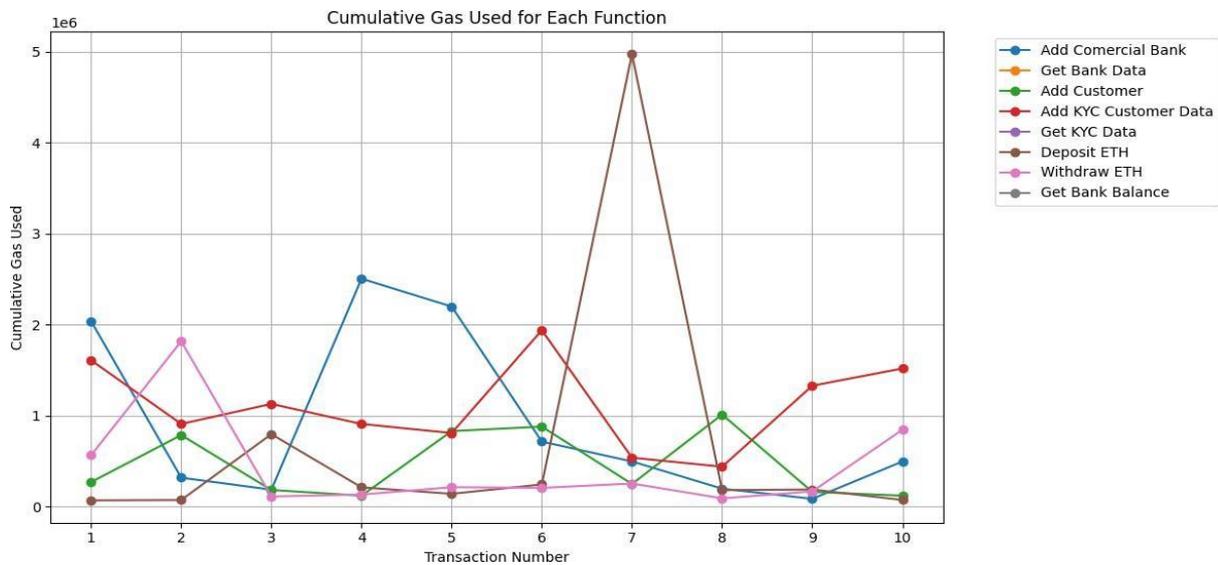

*Figure 17.* Gas Usage of Functions.

## 7. Conclusion

Data security is a critical concern for organizations, institutions, and individuals. A security breach can lead to severe consequences, including data loss, reputational damage, and financial harm. Consequently, it is essential to develop modern decentralized bank applications based on more secure technologies.

Our findings indicate that this approach can enhance the security, reliability, and efficiency of traditional banking systems.

Designed and developed blockchain-based decentralized bank application ALBank leverages blockchain technology and smart contracts to enhance data storage security and operational efficiency. ALBank manages bank transactions and securely stores data on the public Ethereum blockchain. This eliminates risks associated with storing critical data in a centralized database, such as single points of failure, lack of transparency, reliance on untrusted third parties, and vulnerability to attacks.

Furthermore, we developed a module for a decentralized Know Your Customer (KYC) process with autonomous execution using smart contracts, integrating MetaMask for user authentication. This enhances security and addresses the limitations of traditional authentication methods, which rely on storing sensitive credentials like usernames and passwords in centralized databases that are prone to vulnerabilities.

Performance metrics indicate that numerous costly, tedious, and time-consuming banking functions can be efficiently addressed through blockchain-based solutions by reducing costs, improving security, and increasing speed.

Future research directions that should be pursued to further enhance the ALBank include the following:
- Optimization of performance, with particular emphasis on scalability.
- Expanding the functionality of ALBank by adding features such as loans, lending, deposits, checking accounts, mortgage accounts, and investment accounts.
- Development of mechanisms to improve interoperability between different blockchain platforms, thereby enabling a more integrated and flexible ecosystem.
- Identification and mitigation of security vulnerabilities in smart contracts, including the advancement of verification and auditing techniques,
- Ensuring compliance with relevant regulations and legal frameworks.

**References**

[1]. Ali, M. S., Vecchio, M., Pincheira, M., Dolui, K., Antonelli, F., and Rehmani, M. H., "Applications of blockchains in the internet of things: A comprehensive survey," *IEEE Communications Surveys & Tutorials, vol. 21, no. 2, pp. 1676–1717, 2018*.
[2]. Bis, "Distributed ledger technology in payment, clearing and set-tlement. an analytical framework."2017, available online: https://www.bis.org/cpmi/publ/d157.pdf, (Accessed April 2025).
[3]. Buterin, V., "Ethereum whitepaper." 2013, available online: https://ethereum.org/en/whitepaper/, (Accessed April 2025).
[4]. cgaa, "History of banking from ancient civilizations to modern times."2025,available online: https://www.cgaa.org/article/history-of-banking, (Accessed April 2025).
[5]. Chaum, D. L., Computer Systems established, maintained and trusted by mutually suspicious groups. *Electronics Research Laboratory, University of California Riverside, CA, USA, 1979*.
[6]. Di Pierro, M., "What is the blockchain?" *Computing in Science & Engineering, vol. 19, no. 5, pp. 92–95, 2017*.
[7]. Karadag, B., Zaim, A. H., & Akbulut, A. (2024). Blockchain Based KYC Model for Credit Allocation in Banking. *IEEE Access*.
[8]. Lamberty, R., Kirste, D., Kannengießer, N., & Sunyaev, A. (2024). HybCBDC: A Design for Central Bank Digital Currency Systems Enabling Digital Cash. *IEEE Access*.
[9]. Li, W., He, M., and Haiquan, S., "An overview of blockchain technology: applications, challenges and future trends," *11th International Conference on Electronics Information and Emergency Communi- cation (ICEIEC) IEEE, 2021, pp. 31–39*.
[10]. Mainetti, L., Aprile, M., Mele, E., & Vergallo, R. (2023). A sustainable approach to delivering programmable peer-to-peer offline payments. *Sensors, 23(3), 1336*.
[11]. Nakamoto, S., "Bitcoin: a peer-to-peer electronic cash system." 2008, available online: https://bitcoin.org/bitcoin.pdf, (Accessed April 2025).
[12]. Parra Moyano, J. and Ross, O., "Kyc optimization using distributed ledger technology," *Business & Information Systems Engineering, vol. 59, pp. 411–423, 2017*.
[13]. Sakho, S., Jianbiao, Z., Essaf, F., & Badiss, K. (2019, December). Improving banking transactions using blockchain technology. *In 2019 IEEE 5th International Conference on Computer and Communications (ICCC) (pp. 1258-1263). IEEE*.
[14]. Sharma, R., "Bit gold, investopedia, 2024." 1998, available on- line: https://www.investopedia.com/terms/b/bit-gold.asp, (Accessed April 2025).
[15]. Sherifi, S., "Ethereum bank smartcontract kyc," 2025, available online: https://github.com/shkelqimsherifi/Ethereum Bank SmartContract KYC, (Accessed April 2025).

[16]. Sheldon, R., "A timeline and history of blockchain technology." 2008, available online: https://whatis.techtarget.com/feature/A-timeline- and-history-of-blockchain-technology, (Accessed April 2025).

[17]. Szabo, N., "Smart contracts." 1994, available online: https://nakamotoinstitute.org/library/smart-contracts/, (Accessed April 2025).

[18]. Tunzina, T., Chayon, M. A. K., Jitu, P. G., Ankon, M. U., Joy, S. N., Shaha, R. K., ... & Pham, P. H. (2024). Blockchain-based central bank digital currency: Empowering centralized oversight with decentralized transactions. *IEEE Access*.

[19]. usaBank, "Early u.s. banking history." 2025, available online: https://guides.loc.gov/banking-history/banking-history, (Accessed April 2025).

[20]. Wikipedia, "History of banking." 2025, available online: https://en.wikipedia.org/wiki/History of banking, (Accessed April 2025).

[21]. Zheng, Z., Xie, S., Dai, H.-N., Chen, X., and Wang, H., "Blockchain challenges and opportunities: A survey," *International journal of web and grid services, vol. 14, no. 4, pp. 352– 375, 2018*.